\documentclass[journal,onecolumn,12pt,twoside,a4paper]{IEEEtran}

\usepackage{amsmath}
\usepackage{cite}
\usepackage{latexsym}
\usepackage{graphics}
\usepackage{amssymb}
\usepackage{amsthm}
\usepackage{url}
\usepackage{cite}
\usepackage{array}
\usepackage{graphicx}
\usepackage{caption}
\usepackage{subcaption}

\newcommand{\maximize}[1]{{\underset{{#1}}{\mathrm{maximize}}}}
\newcommand{\minimize}[1]{{\underset{{#1}}{\mathrm{minimize}}}}
\newcommand{\fracSum}[1]{{\underset{{#1}}{\sum}}}
\newcommand{\fracSumtwo}[2]{\overset{#2}{\underset{#1}{\sum}}}
\newcommand{\argmax}[1]{{\underset{{#1}}{\mathrm{arg\,max}}}}

\newcommand{\vect}[1]{\mathbf{#1}}
\def\diag{\mathrm{diag}}

\begin{document}

\title{Optimal Multiuser Transmit Beamforming: \\A~Difficult~Problem~with~a~Simple~Solution~Structure}

\author{Emil Bj\"ornson, Mats Bengtsson, and Bj\"orn Ottersten\thanks{\copyright 2014 IEEE. Personal use of this material is permitted. Permission from IEEE must be obtained for all other uses, in any current or future media, including reprinting/republishing this material for advertising or promotional purposes, creating new collective works, for resale or redistribution to servers or lists, or reuse of any copyrighted component of this work in other works. \newline \indent This lecture note has been accepted for publication in IEEE Signal Processing Magazine. \newline \indent
Supplementary downloadable material is available at https://github.com/emilbjornson/optimal-beamforming, provided by the authors. The material includes Matlab code that can reproduce all simulation results. Contact emil.bjornson@liu.se for further questions about this work.}}

\maketitle

\IEEEpeerreviewmaketitle

\section*{}
\vspace{-1.5cm}

Transmit beamforming is a versatile technique for signal transmission from an array of $N$ antennas to one or multiple users \cite{Gershman2010a}. In wireless communications, the goal is to increase the signal power at the intended user and reduce interference to non-intended users. A high signal power is achieved by transmitting the same data signal from all antennas, but with different amplitudes and phases, such that the signal components add \emph{coherently} at the user. Low interference is accomplished by making the signal components add \emph{destructively} at non-intended users.
This corresponds mathematically to designing beamforming vectors (that describe the amplitudes and phases) to have large inner products with the vectors describing the intended channels and small inner products with non-intended user channels.

\begin{figure}[t!]
\begin{center}
\includegraphics[width=10cm]{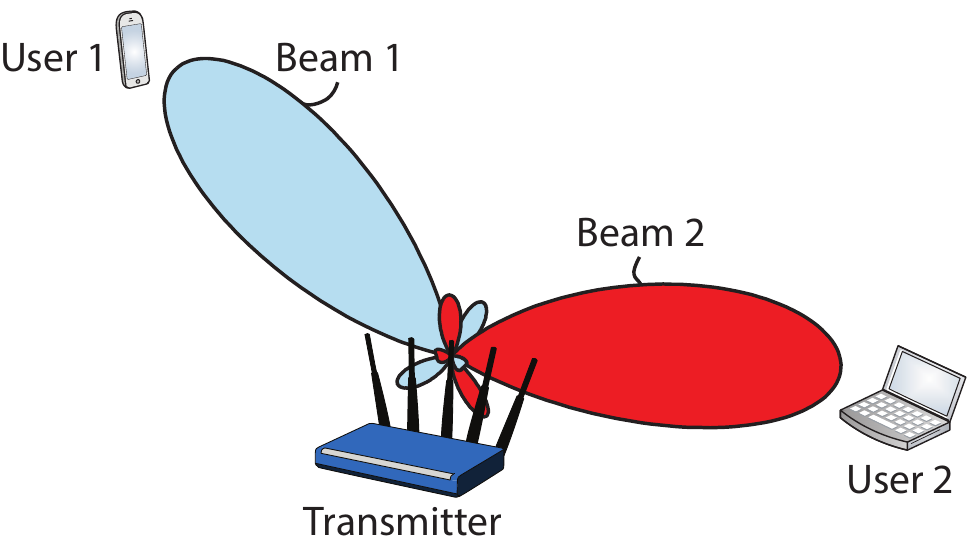}
\end{center} \vskip -2mm
\caption{Visualization of transmit beamforming in an LoS scenario. The beamforming is adapted to the location of the intended user, such that a main-lobe with a strong signal power is achieved toward this user while the side-lobes that cause interference to other non-intended users are weak.}\label{figure_beamforming-illustration} \vskip-1mm
\end{figure}

If there is line-of-sight (LoS) between the transmitter and receiver, beamforming can be seen as forming a signal beam toward the receiver; see Figure \ref{figure_beamforming-illustration}. Beamforming can also be applied in non-LoS scenarios, if the multipath channel is known, by making the multipath components add coherently or destructively.

Since transmit beamforming focuses the signal energy at certain places, less energy arrives to other places.
This allows for so-called space-division multiple access (SDMA), where $K$ spatially separated users are served simultaneously. One beamforming vector is assigned to each user and can be matched to its channel. Unfortunately, the finite number of transmit antennas only provides a limited amount of spatial directivity, which means that there are energy leakages between the users which act as interference.

While it is fairly easy to design a beamforming vector that maximizes the signal power at the intended user, it is difficult to strike a perfect balance between maximizing the signal power and minimizing the interference leakage. In fact, the optimization of multiuser transmit beamforming is generally a nondeterministic polynomial-time (NP) hard problem \cite{Liu2011a}. Nevertheless, this lecture shows that the optimal transmit beamforming has a simple structure with very intuitive properties and interpretations. This structure provides a theoretical foundation for practical low-complexity beamforming schemes.

\section*{Relevance}

Adaptive transmit beamforming is key to increased spectral and energy efficiency in next-generation wireless networks, which are expected to include very large antenna arrays \cite{Rusek2013a}. In light of the difficulty to compute the optimal multiuser transmit beamforming, there is a plethora of heuristic schemes. Although each scheme might be optimal in some special case, and can be tweaked to fit other cases, these heuristic schemes generally do not provide sufficient degrees of freedom to ever achieve the optimal performance. The main purpose of this lecture is to provide a structure of optimal linear transmit beamforming, with a sufficient number of design parameters to not lose optimality. This simple structure provides many insights and is easily extended to take various design constraints of practical cellular networks into account.

\section*{Prerequisites}

The readers require basic knowledge in linear algebra, communication theory, and convex optimization.

\section*{Problem (P1): Power Minimization with SINR Constraints}

We consider a downlink channel where a base station (BS) equipped with $N$ antennas communicates with $K$ single-antenna users using SDMA. The data signal to user $k$ is denoted $s_k \in \mathbb{C}$ and is normalized to unit power, while the vector $\vect{h}_k \in \mathbb{C}^{N \times 1}$ describes the corresponding channel. The $K$ different data signals are separated spatially using the linear beamforming vectors $\vect{w}_1,\ldots, \vect{w}_K \in \mathbb{C}^{N \times 1}$, where $\vect{w}_k$ is associated with user $k$. The normalized version $\frac{\vect{w}_k}{\| \vect{w}_k\|}$ is called the beamforming direction and points out a direction in the $N$-dimensional vector space---note that it only corresponds to a physical direction in LoS scenarios. The squared norm $\| \vect{w}_k \|^2$ is the power allocated for transmission to user $k$.
We model the received signal $r_k \in \mathbb{C}$ at user $k$ as
\begin{equation}
r_k =  \vect{h}_k^H \left( \sum_{i=1}^{K} \vect{w}_i s_i \right) + n_k
\end{equation}
where $n_k$ is additive receiver noise with zero mean and variance $\sigma^2$. Consequently, the signal-to-noise-and-interference ratio (SINR) at user $k$ is
\begin{align}
\mathrm{SINR}_k = \frac{| \vect{h}_k^H \vect{w}_k|^2 }{ \fracSum{i \neq k} | \vect{h}_k^H \vect{w}_i|^2 + \sigma^2} = \frac{ \frac{1}{\sigma^2} | \vect{h}_k^H \vect{w}_k|^2 }{  \fracSum{i \neq k} \frac{1}{\sigma^2} | \vect{h}_k^H \vect{w}_i|^2 + 1}. \label{eq:SINRs}
\end{align}
We use the latter, noise-normalized expression in this lecture since it emphasizes the impact of the noise.

The transmit beamforming can be optimized to maximize some performance utility metric, which is generally a function of the SINRs. The main goal of this lecture is to analyze a general formulation of such a problem, defined later as Problem \eqref{eq:generalized-problem}, and to derive the structure of the optimal beamforming. As a preparation toward this goal, we first solve the relatively simple power minimization problem
\begin{equation} \label{eq:original-problem} \tag{P1}
\begin{split}
\minimize{\vect{w}_1,\ldots,\vect{w}_{K}}\, & \quad \sum_{k=1}^{K} \| \vect{w}_k \|^2 \\
\mathrm{subject} \,\, \mathrm{to} & \quad \mathrm{SINR}_k \geq \gamma_k.
\end{split}
\end{equation}
The parameters $\gamma_1,\ldots,\gamma_K$ are the SINRs that each user shall achieve at the optimum of \eqref{eq:original-problem}, using as little transmit power as possible. The $\gamma$-parameters can, for example, describe the SINRs required for achieving certain data rates. The values of the $\gamma$-parameters are constant in (P1) and clearly impact the optimal beamforming solution, but we will see later that the solution structure is always the same.

\section*{Solution to Problem (P1)}

The first step toward solving \eqref{eq:original-problem} is to reformulate it as a convex problem. The cost function $\sum_{k=1}^{K} \| \vect{w}_k \|^2$ is clearly a convex function of the beamforming vectors. To extract the hidden convexity of the SINR constraints, $\mathrm{SINR}_k \geq \gamma_k$, we make use of a trick from \cite{Bengtsson2001a}. We note that the absolute values in the SINRs in \eqref{eq:SINRs} make $\vect{w}_k$ and $e^{\jmath \theta_k} \vect{w}_k$ completely equivalent for any common phase rotation $\theta_k \in \mathbb{R}$. Without loss of optimality, we exploit this phase ambiguity to rotate the phase such that the inner product $\vect{h}_k^H \vect{w}_k$ is real-valued and positive. This implies that $\sqrt{| \vect{h}_k^H \vect{w}_k|^2} = \vect{h}_k^H \vect{w}_k \geq 0$. By letting $\Re (\cdot)$ denoting the real part, the constraint $\mathrm{SINR}_k \geq \gamma_k$ can be rewritten as
\begin{align}
& \frac{1}{\gamma_k \sigma^2} | \vect{h}_k^H \vect{w}_k|^2 \geq   \fracSum{i \neq k} \frac{1}{\sigma^2} | \vect{h}_k^H \vect{w}_i|^2 + 1
 \Leftrightarrow \frac{1}{\sqrt{\gamma_k \sigma^2}}  \Re ( \vect{h}_k^H \vect{w}_k ) \geq  \sqrt{ \fracSum{i \neq k} \frac{1}{\sigma^2} | \vect{h}_k^H \vect{w}_i|^2 + 1 }. \label{eq:socp-constraint}
\end{align}
The reformulated SINR constraint in \eqref{eq:socp-constraint} is a second-order cone constraint, which is a convex type of constraint \cite{Bengtsson2001a,Wiesel2006a,Yu2007a}, and it is easy to show that Slater's constraint qualification is fulfilled \cite{Bjornson2013d}.
Hence, optimization theory provides many important properties for the reformulated convex problem; in particular, strong duality and that the Karush-Kuhn-Tucker (KKT) conditions are necessary and sufficient for the optimal solution. It is shown in \cite[Appendix A]{Yu2007a} (by a simple parameter change) that these properties also hold for the original problem \eqref{eq:original-problem}, although \eqref{eq:original-problem} is not convex. The strong duality and KKT conditions for \eqref{eq:original-problem} play a key role in this lecture. To show this, we define the Lagrangian function of \eqref{eq:original-problem} as
\begin{equation} \label{eq_lagrangian}
\mathcal{L}(\vect{w}_1,\ldots,\vect{w}_{K},\lambda_1,\ldots,\lambda_K) = \sum_{k=1}^{K} \| \vect{w}_k \|^2 + \sum_{k=1}^{K} \lambda_k \left( \fracSum{i \neq k} \frac{1}{\sigma^2} | \vect{h}_k^H \vect{w}_i|^2 + 1 - \frac{1}{\gamma_k \sigma^2} | \vect{h}_k^H \vect{w}_k|^2 \right)
\end{equation}
where $\lambda_k \geq 0$ is the Lagrange multiplier associated with the $k$th SINR constraint. The dual function is $\min_{\vect{w}_1,\ldots,\vect{w}_{K}} \mathcal{L} = \sum_{k=1}^{K} \lambda_k$ and the strong duality implies that it equals the total power $\sum_{k=1}^{K} \| \vect{w}_k \|^2$ at the optimal solution, which we utilize later in this lecture.
To solve \eqref{eq:original-problem}, we now exploit the stationarity KKT conditions which say that $\partial \mathcal{L}/\partial \vect{w}_k =
\vect{0}$, for $k=1,\ldots,K$, at the optimal solution. This implies
\begin{align}  \label{eq:stationarity-condition-part1}
 &\vect{w}_k  + \fracSum{i \neq k}  \frac{\lambda_i}{\sigma^2} \vect{h}_i \vect{h}_i^H \vect{w}_k - \frac{\lambda_k}{\gamma_k \sigma^2}\vect{h}_k \vect{h}_k^H \vect{w}_k = \vect{0} \\ \label{eq:stationarity-condition-part2}
 & \quad \Leftrightarrow  \quad \left( \vect{I}_N +  \sum_{i=1}^{K} \frac{\lambda_i}{\sigma^2} \vect{h}_i \vect{h}_i^H \right) \vect{w}_k = \frac{\lambda_k}{\sigma^2} \left(1+\frac{1}{\gamma_k}\right) \vect{h}_k  \vect{h}_k^H \vect{w}_k\\
 & \quad \Leftrightarrow  \quad  \vect{w}_k = \left( \vect{I}_N + \sum_{i=1}^{K} \frac{\lambda_i}{\sigma^2} \vect{h}_i \vect{h}_i^H \right)^{-1} \vect{h}_k \underbrace{ \frac{\lambda_k}{\sigma^2} \left(1+\frac{1}{\gamma_k}\right) \vect{h}_k^H \vect{w}_k }_{=\textrm{scalar}} \label{eq:stationarity-condition}
\end{align}
where $\vect{I}_N$ denotes the $N \times N$ identity matrix. The expression \eqref{eq:stationarity-condition-part2} is achieved from \eqref{eq:stationarity-condition-part1} by adding the term $\frac{\lambda_k}{\sigma^2} \vect{h}_k \vect{h}_k^H \vect{w}_k$ to both sides and \eqref{eq:stationarity-condition} is obtained by multiplying with an inverse. Since $(\lambda_k/\sigma^2) (1+1/\gamma_k) \vect{h}_k^H \vect{w}_k$ is a scalar, \eqref{eq:stationarity-condition} shows that the optimal $\vect{w}_k$ must be parallel to $( \vect{I}_N +  \sum_{i =1}^{K} \frac{\lambda_i}{\sigma^2} \vect{h}_i \vect{h}_i^H )^{-1} \vect{h}_k$. In other words, the optimal beamforming vectors $\vect{w}_1^\star,\ldots,\vect{w}_{K}^\star$ are
\begin{equation} \label{eq:beamforming-solution}
\vect{w}_k^\star = \underbrace{\sqrt{p_k}}_{=\textrm{beamforming power}} \underbrace{\frac{\Big( \vect{I}_N +  \fracSumtwo{i=1}{K} \frac{\lambda_i}{\sigma^2} \vect{h}_i \vect{h}_i^H \Big)^{-1} \vect{h}_k }{ \bigg\| \Big( \vect{I}_N +  \fracSumtwo{i=1}{K} \frac{\lambda_i}{\sigma^2} \vect{h}_i \vect{h}_i^H \Big)^{-1} \vect{h}_k  \bigg\|}}_{=\tilde{\vect{w}}_k^{\star} = \textrm{beamforming direction}} \qquad \mathrm{for} \,\,\, k=1,\ldots,K
\end{equation}
where $p_k$ denotes the beamforming power and $\tilde{\vect{w}}_k^{\star}$ denotes the unit-norm beamforming direction for user $k$. The $K$ unknown beamforming powers are computed by noting that the SINR constraints \eqref{eq:socp-constraint} hold with equality at the optimal solution. This implies $\frac{1}{\gamma_k} p_k | \vect{h}_k^H \tilde{\vect{w}}_k^{\star}|^2 -  \sum_{i \neq k} p_i | \vect{h}_k^H \tilde{\vect{w}}_i^{\star}|^2 = \sigma^2$ for $k=1,\ldots,K$. Since we know the beamforming directions, we have $K$ linear equations and obtain the $K$ powers as
\begin{equation} \label{eq:beamforming-power}
\begin{bmatrix}
        p_1  \\[-2mm]
       \vdots \\[-2mm]
        p_K
      \end{bmatrix} =  \vect{M}^{-1}  \begin{bmatrix}
        \sigma^2  \\[-2mm]
       \vdots \\[-2mm]
        \sigma^2
      \end{bmatrix}  \qquad \mathrm{where} \quad [ \vect{M} ]_{ij} = \begin{cases} \frac{1}{\gamma_i} | \vect{h}_{i}^H
\tilde{\vect{w}}_i^{\star}|^2, & i = j, \\ - | \vect{h}_{i}^H \tilde{\vect{w}}_j^{\star}|^2, & i \neq j, \end{cases}
\end{equation}
and $[ \vect{M} ]_{ij}$ denotes the $(i,j)$th element of the matrix $\vect{M} \in \mathbb{R}^{K \times K}$.

By combining \eqref{eq:beamforming-solution} and \eqref{eq:beamforming-power}, we obtain the structure of optimal beamforming as a function of the Lagrange multipliers $\lambda_1,\ldots,\lambda_K$. Finding these multipliers is outside the scope of this lecture, for reasons that will be clear in the next section. However, we note that the Lagrange multipliers can be computed by convex optimization \cite{Bengtsson2001a} or from the fixed-point equations $\lambda_k = \frac{\sigma^2}{( 1+ \frac{1}{\gamma_k}) \vect{h}_k^H ( \vect{I}_N +  \sum_{i=1}^{K} \frac{\lambda_i}{\sigma^2} \vect{h}_i \vect{h}_i^H )^{-1} \vect{h}_k}$  for all $k$ \cite{Wiesel2006a,Yu2007a}.

\section*{Problem (P2): General Transmit Beamforming Optimization}

The main goal of this lecture is to analyze a very general transmit beamforming optimization problem. We want to maximize some arbitrary utility function $f(\mathrm{SINR}_1, \ldots, \mathrm{SINR}_K)$ that is strictly increasing in the SINR of each user, while the total transmit power is limited by $P$.  This is stated mathematically as
\begin{equation} \label{eq:generalized-problem} \tag{P2}
\begin{split}
\maximize{\vect{w}_1,\ldots,\vect{w}_{K}}\, & \quad f( \mathrm{SINR}_1, \ldots, \mathrm{SINR}_K) \\
\mathrm{subject} \,\, \mathrm{to} & \quad \sum_{k=1}^{K} \| \vect{w}_k \|^2 \leq P.
\end{split}
\end{equation}

Despite the conciseness of \eqref{eq:generalized-problem}, it is generally very hard to solve \cite{Bjornson2013d}. Indeed, \cite{Liu2011a} proves that it is NP-hard for many common utility functions; for example, the sum rate $f( \mathrm{SINR}_1, \ldots, \mathrm{SINR}_K) = \sum_{k=1}^K \log_2(1+\mathrm{SINR}_k)$. Nevertheless, we will show that the structure of the optimal solution to \eqref{eq:generalized-problem} is easily obtained.

\section*{Solution Structure to Problem (P2)}

Suppose for the moment that we know the SINR values $\mathrm{SINR}_1^\star, \ldots, \mathrm{SINR}_K^\star$ that are achieved by the optimal solution to \eqref{eq:generalized-problem}. What would happen if we set $\gamma_k = \mathrm{SINR}_k^\star$, for $k=1,\ldots,K$, and solve \eqref{eq:original-problem} for these particular $\gamma$-parameters? The answer is that the beamforming vectors that solve \eqref{eq:original-problem} will now also solve \eqref{eq:generalized-problem} \cite{Bjornson2013d}. This is understood as follows:
\eqref{eq:original-problem} finds beamforming vectors that achieves the SINR values $\mathrm{SINR}_1^\star, \ldots, \mathrm{SINR}_K^\star$. The solution to \eqref{eq:original-problem} must satisfy the total power constraint in \eqref{eq:generalized-problem}, because \eqref{eq:original-problem} gives the beamforming that achieves the given SINRs using the minimal amount of power. Since the beamforming vectors from \eqref{eq:original-problem} are feasible for \eqref{eq:generalized-problem} and achieves the optimal SINR values, they are an optimal solution to \eqref{eq:generalized-problem} as well.

We can, of course, not know $\mathrm{SINR}_1^\star, \ldots, \mathrm{SINR}_K^\star$ unless we actually solve \eqref{eq:generalized-problem}. In fact, the difference between the relatively easy \eqref{eq:original-problem} and the difficult \eqref{eq:generalized-problem} is that the SINRs are predefined in \eqref{eq:original-problem} while we need to find the optimal SINR values (along with the beamforming vectors) in \eqref{eq:generalized-problem}. The connection between the two problem implies, however, that the optimal beamforming for \eqref{eq:generalized-problem} is
\begin{equation} \label{eq:beamforming-solution2}
\vect{w}_k^\star = \sqrt{p_k} \frac{\Big( \vect{I}_N +  \fracSumtwo{i=1}{K} \frac{\lambda_i}{\sigma^2} \vect{h}_i \vect{h}_i^H \Big)^{-1} \vect{h}_k }{ \bigg\| \Big( \vect{I}_N +  \fracSumtwo{i=1}{K} \frac{\lambda_i}{\sigma^2} \vect{h}_i \vect{h}_i^H \Big)^{-1} \vect{h}_k  \bigg\|} \qquad \mathrm{for} \,\,\, k=1,\ldots,K
\end{equation}
for some positive parameters $\lambda_1,\ldots,\lambda_K$. The strong duality property of \eqref{eq:original-problem} implies $\sum_{i=1}^{K} \lambda_i = P$, since
$P$ is the optimal cost function in \eqref{eq:original-problem} and $\sum_{i=1}^{K} \lambda_i$ is the dual function.
Finding the optimal parameter values in this range is equivalent to solving \eqref{eq:generalized-problem}, thus it is as hard as solving the original problem. However, the importance of \eqref{eq:beamforming-solution2} is that it provides a simple structure for the optimal beamforming.

Since the matrix inverse in \eqref{eq:beamforming-solution2} is the same for all users, the matrix $\vect{W}^\star = [\vect{w}_1^\star \, \ldots \, \vect{w}_K^\star] \in \mathbb{C}^{N \times K}$ with the optimal beamforming vectors can be written in a compact form. To this end, we note that $\sum_{i=1}^{K} \frac{\lambda_i}{\sigma^2} \vect{h}_i \vect{h}_i^H = \frac{1}{\sigma^2} \vect{H} \vect{\Lambda} \vect{H}^H$ where $\vect{H} = [\vect{h}_1 \, \ldots \, \vect{h}_K] \in \mathbb{C}^{N \times K}$ contains the channels and $\vect{\Lambda} = \diag( \lambda_1, \ldots, \lambda_K)$ is a diagonal matrix with the $\lambda$-parameters. By gathering the power allocation in a matrix $\vect{P}$, we obtain
\begin{equation}  \label{eq:beamforming-solution_matrixform}
\vect{W}^\star  = \left( \vect{I}_N + \frac{1}{\sigma^2} \vect{H} \vect{\Lambda} \vect{H}^H \right)^{-1} \vect{H} \vect{P}^{\frac{1}{2}}
\end{equation}
where $\vect{P} = \diag( p_1 / \| ( \vect{I}_N +  \frac{1}{\sigma^2} \vect{H} \vect{\Lambda} \vect{H}^H )^{-1} \vect{h}_1  \|^2, \ldots, p_K / \| ( \vect{I}_N +  \frac{1}{\sigma^2} \vect{H} \vect{\Lambda} \vect{H}^H )^{-1} \vect{h}_K  \|^2)$ and $(\cdot)^{\frac{1}{2}}$ denotes the matrix square root. In the next section we study the structure of \eqref{eq:beamforming-solution2} and \eqref{eq:beamforming-solution_matrixform} and gain some insights.

\section*{Intuition Behind the Optimal Structure}

\begin{figure}[t!]
\begin{center}
\includegraphics[width=10cm]{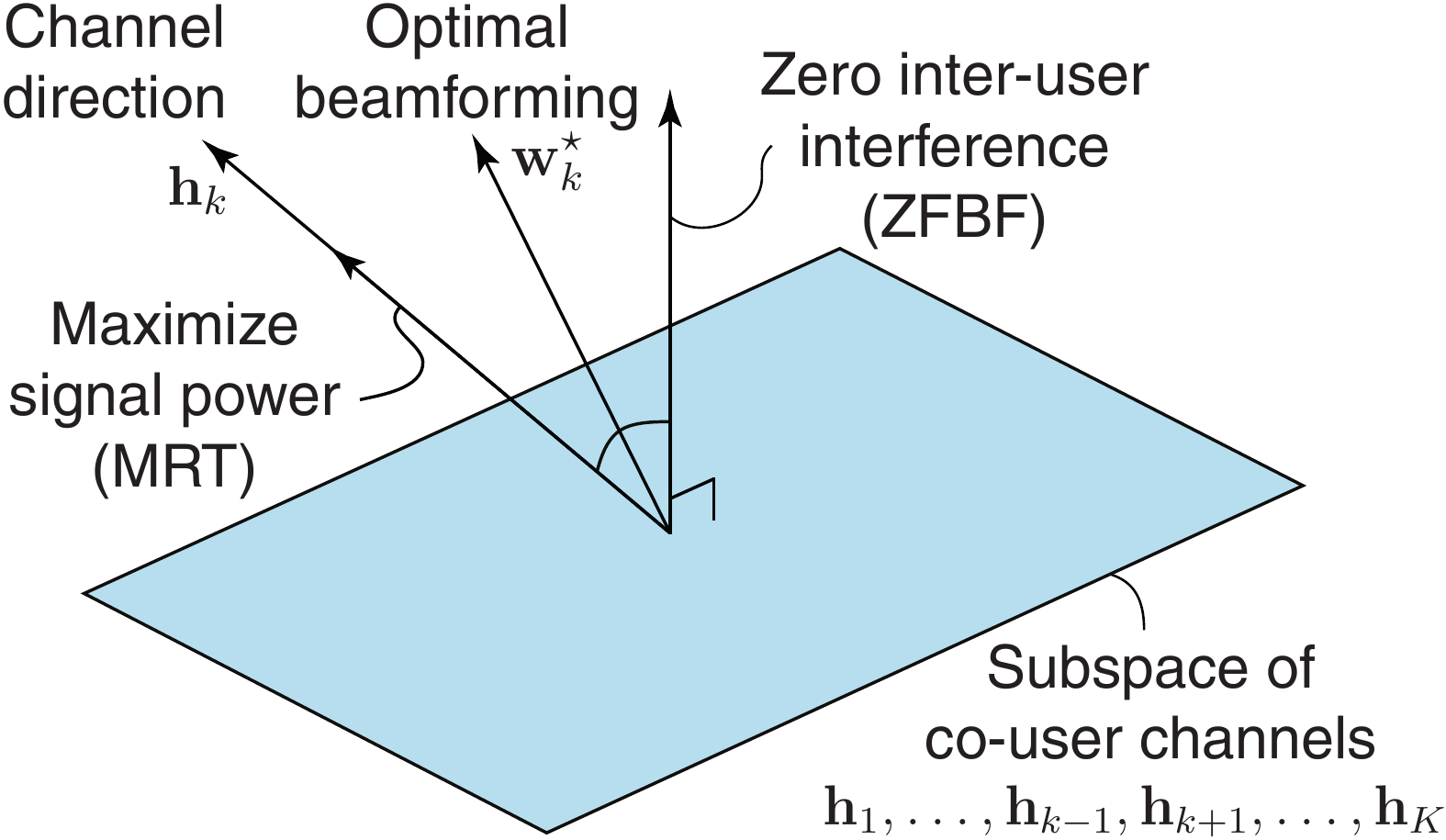}
\end{center} \vskip -2mm
\caption{The optimal beamforming $\vect{w}_k^\star$ is based on the channel direction $\vect{h}_k$ but rotated to balance between high signal power and being orthogonal to the co-user channels.}\label{figure_geometric-illustration} \vskip-1mm
\end{figure}

The optimal beamforming direction in \eqref{eq:beamforming-solution2} consists of two main parts: 1) the channel vector $\vect{h}_k$ between the BS and the intended user $k$; and 2) the matrix $( \vect{I}_N +  \sum_{i=1}^{K} \frac{\lambda_i}{\sigma^2} \vect{h}_i \vect{h}_i^H )^{-1}$. Beamforming in the same direction as the channel (i.e., $\tilde{\vect{w}}_k^{\mathrm{(MRT)}} = \frac{\vect{h}_k}{\| \vect{h}_k \|}$) is known as \emph{maximum ratio transmission} (MRT) or matched filtering \cite{Lo1999a}. This selection maximizes the received signal power $p_k | \vect{h}_k^H \tilde{\vect{w}}_k|^2$ at the intended user, because
\begin{equation}
\argmax{\tilde{\vect{w}}_k : \, \| \tilde{\vect{w}}_k \|^2 = 1 } \quad | \vect{h}_k^H \tilde{\vect{w}}_k|^2 = \frac{\vect{h}_k}{\| \vect{h}_k \|}
\end{equation}
due to the Cauchy-Schwarz inequality. This is the optimal beamforming direction for $K=1$, but not when there are multiple users because the inter-user interference is unaccounted for in MRT. This is basically what the multiplication of $\vect{h}_k$ with $( \vect{I}_N +  \sum_{i=1}^{K} \frac{\lambda_i}{\sigma^2} \vect{h}_i \vect{h}_i^H )^{-1}$ (before normalization) takes care of; it rotates MRT to reduce the interference that is caused in the co-user directions $\vect{h}_1,\ldots,\vect{h}_{k-1},\vect{h}_{k+1},\ldots,\vect{h}_K$. This interpretation is illustrated in Figure \ref{figure_geometric-illustration}, where the optimal beamforming lies somewhere in between MRT and the vector that is orthogonal to all co-user channels.
The optimal beamforming direction depends ultimately on the utility function $f(\cdot,\ldots,\cdot)$. However, the parameter $\lambda_i \geq 0$ can be seen as the priority of user $i$, where a larger value means that other users' beamforming vectors will be more orthogonal to $\vect{h}_i$.

\subsection*{Asymptotic Properties}

Next, we study the asymptotic beamforming properties.
In the low signal-to-noise ratio (SNR) case, represented by $\sigma^2 \rightarrow \infty$, the system is noise-limited and the beamforming matrix in \eqref{eq:beamforming-solution_matrixform} converges to
\begin{equation}  \label{eq:beamforming-lowSNR}
\vect{W}^\star_{\sigma^2 \rightarrow \infty}  = (\vect{I}_N + \vect{0})^{-1} \vect{H} \vect{P}^{\frac{1}{2}}_{\sigma^2 \rightarrow \infty} = \vect{H} \vect{P}^{\frac{1}{2}}_{\sigma^2 \rightarrow \infty}
\end{equation}
where the matrix inverse vanishes and $\vect{P}_{\sigma^2 \rightarrow \infty}$ denotes the asymptotic power allocation. This implies that $\vect{w}_k^\star$ is a scaled version of the channel vector $\vect{h}_k$, which is equivalent to MRT.

At high SNRs, given by $\sigma^2 \rightarrow 0$, the system is interference-limited. We focus on the case $N \geq K$ with at least one spatial degree-of-freedom per user---this is the meaningful operating regime for SDMA. To avoid singularity in the inverse when $\sigma^2$ is small, we use the identity $(\vect{I} + \vect{A} \vect{B})^{-1} \vect{A} = \vect{A} (\vect{I} + \vect{B} \vect{A})^{-1}$ and rewrite \eqref{eq:beamforming-solution_matrixform} as
$\vect{W}^\star  = \vect{H} \left( \sigma^2 \vect{I}_K + \vect{\Lambda} \vect{H}^H \vect{H} \right)^{-1} \tilde{\vect{P}}^{\frac{1}{2}}$
where $\tilde{\vect{P}} = \diag( p_1 / \| ( \sigma^2 \vect{I}_K + \vect{\Lambda} \vect{H}^H \vect{H} )^{-1} \vect{h}_1  \|^2, \ldots, p_K / \| ( \sigma^2 \vect{I}_K + \vect{\Lambda} \vect{H}^H \vect{H} )^{-1} \vect{h}_K  \|^2)$ denotes the corresponding rewritten power allocation matrix. It now follows that
\begin{equation} \label{eq:beamforming-highSNR}
\vect{W}^\star_{\sigma^2 \rightarrow 0}  =
\vect{H} \left( 0 \vect{I}_K + \vect{\Lambda} \vect{H}^H \vect{H} \right)^{-1} \tilde{\vect{P}}_{\sigma^2 \rightarrow 0}
= \vect{H} \left( \vect{H}^H \vect{H} \right)^{-1} \vect{\Lambda}^{-1} \tilde{\vect{P}}_{\sigma^2 \rightarrow 0}
\end{equation}
where the term $\sigma^2 \vect{I}_K$ vanishes when $\sigma^2 \rightarrow 0$ and $\tilde{\vect{P}}_{\sigma^2 \rightarrow 0}$ denotes the asymptotic power allocation. This solution is known as channel inversion or \emph{zero-forcing beamforming} (ZFBF) \cite{Joham2005a}, because it contains the pseudo-inverse $\vect{H} \left( \vect{H}^H \vect{H} \right)^{-1}$ of the channel matrix $\vect{H}^H$. Hence, $\vect{H}^H \vect{W}^\star_{\sigma^2 \rightarrow 0} = \vect{\Lambda}^{-1} \tilde{\vect{P}}_{\sigma^2 \rightarrow 0}$ is a diagonal matrix. Since the off-diagonal elements are of the form $\vect{h}_i^H \vect{w}_k^\star = 0$ for $i \neq k$, this beamforming causes zero inter-user interference by projecting $\vect{h}_k$ onto the subspace that is orthogonal to the co-user channels.

The asymptotic properties are intuitive if we look at the SINR in \eqref{eq:SINRs}. The noise dominates over the interference at low SNRs, thus we should use MRT to maximize the signal power without caring about interference. On the contrary, the interference dominates over the noise at high SNRs, thus we should use ZFBF to remove it. We recall from Figure~\ref{figure_geometric-illustration} that MRT and ZFBF are also the two extremes from a geometric perspective and the optimal beamforming at arbitrary SNR balance between these extremes.

Another asymptotic regime has received much attention: the use of very large arrays where the number of antennas, $N$, goes to infinity in the performance analysis \cite{Rusek2013a}. A key motivation is that the squared channel norms ($\| \vect{h}_k \|^2$) are proportional to $N$, while the cross-products ($|\vect{h}_i^H \vect{h}_k |$ for $i \neq k$) increase more slowly with $N$ (the exact scaling depends on the channel models). Hence, the user channels become orthogonal as $N \rightarrow \infty$, which reduces interference and allows for less transmit power. Observe that $\sigma^2 \vect{I}_K + \vect{\Lambda} \vect{H}^H \vect{H} \approx \vect{\Lambda} \vect{H}^H \vect{H}$ for large $N$, since only the elements of $\vect{H}^H \vect{H}$ grow with $N$. Similar to \eqref{eq:beamforming-highSNR}, one can then prove that ZFBF is asymptotically optimal. MRT performs relatively well in this regime due to the asymptotic channel orthogonality, but will \emph{not} reach the same performance as ZFBF \cite[Table 1]{Rusek2013a}.

\subsection*{Relationship to Receive Beamforming}

There are striking similarities between transmit beamforming in the downlink and receive beamforming in the uplink, but also fundamental differences. To describe these, we consider the uplink scenario where the same $K$ users are transmitting to the same BS. The received signal $\vect{r} \in \mathbb{C}^{N \times 1}$ at the BS is $\vect{r} = \sum_{i=1}^{K} \vect{h}_i s_i + \vect{n}$, where user $k$ transmits the data signal $s_k$ using the uplink transmit power $q_k$. The receiver noise $\vect{n}$ has zero mean and the covariance matrix $\sigma^2 \vect{I}_N$. The uplink SINR for the signal from user $k$ is
\begin{align}
\mathrm{SINR}_k^{\mathrm{uplink}} =  \frac{ \frac{q_k}{\sigma^2} | \vect{h}_k^H \vect{v}_k|^2 }{  \fracSum{i \neq k} \frac{q_i}{\sigma^2} | \vect{h}_i^H \vect{v}_k|^2 +  \vect{v}_k^H \vect{I}_N \vect{v}_k } \label{eq:SINRs_uplink}
\end{align}
where $\vect{v}_k \in \mathbb{C}^{N \times 1}$ is the unit-norm receive beamforming vector used by the BS to spatially discriminate the signal sent by user $k$ from the interfering signals. The uplink SINR in \eqref{eq:SINRs_uplink} is similar to the downlink SINR in \eqref{eq:SINRs}, but the noise term is scaled by $\| \vect{v}_k \|^2$ and the indices are swapped in the interference term: $| \vect{h}_k^H \vect{w}_i|^2$ in the downlink is replaced by $q_i | \vect{h}_i^H \vect{v}_k|^2$ in the uplink. The latter is because downlink interference originates from the beamforming vectors of other users, while uplink interference arrives through the channels from other users. This tiny difference has a fundamental impact on the optimization, because
the uplink SINR of user $k$ only contains its own receive beamforming vector $\vect{v}_k$. We can therefore optimize the beamforming separately for each $k$:
\begin{equation} \label{eq:uplink-problem}
\argmax{\vect{v}_k : \, \| \vect{v}_k \|^2 = 1 } \quad \frac{ \frac{q_k}{\sigma^2} | \vect{h}_k^H \vect{v}_k|^2 }{ \fracSum{i \neq k} \frac{q_i}{\sigma^2} | \vect{h}_i^H \vect{v}_k|^2 + \vect{v}_k^H \vect{I}_N \vect{v}_k } = \frac{\Big( \vect{I}_N +  \fracSumtwo{i=1}{K} \frac{q_i}{\sigma^2} \vect{h}_i \vect{h}_i^H \Big)^{-1} \vect{h}_k }{ \bigg\| \Big( \vect{I}_N +  \fracSumtwo{i=1}{K} \frac{q_i}{\sigma^2} \vect{h}_i \vect{h}_i^H \Big)^{-1} \vect{h}_k  \bigg\|}.
\end{equation}
The solution follows since this is the maximization of a generalized Rayleigh quotient \cite{Bjornson2013d}. Note that the same receive beamforming is optimal irrespective of which function of the uplink SINRs we want to optimize. In fact, \eqref{eq:uplink-problem} also minimizes the mean squared error (MSE) between the transmitted signal and the processed received signal, thus it is known as the Wiener filter and minimum MSE (MMSE) filter \cite{Joham2005a}.

The optimal transmit and receive beamforming have the same structure; the Wiener filter in \eqref{eq:uplink-problem} is obtained from \eqref{eq:beamforming-solution2} by setting $\lambda_k$ equal to the uplink transmit power $q_k$. This parameter choice is only optimal for the downlink in symmetric scenarios, as discussed later. In general, the parameters are different because the uplink signals pass through different channels (thus, the uplink is affected by variations in the channel norms), while everything that reaches a user in the downlink has passed through a single channel \cite{Bengtsson2001a}.

\subsection*{Heuristic Transmit Beamforming}

It is generally hard to find the optimal $\lambda$-parameters, but the beamforming structure in \eqref{eq:beamforming-solution2} and \eqref{eq:beamforming-solution_matrixform} serves as a foundation for heuristic beamforming; that is, we can select the parameters judiciously and hope for close-to-optimal beamforming.
If we make all the parameters equal, $\lambda_k = \lambda$ for all $k$, we obtain
\begin{equation} \label{eq:beamforming-RZF}
\vect{W}  = \left( \vect{I}_N + \frac{\lambda}{\sigma^2} \vect{H}  \vect{H}^H \right)^{-1} \vect{H} \vect{P}^{\frac{1}{2}} = \vect{H}  \left( \vect{I}_K + \frac{\lambda}{\sigma^2}\vect{H}^H \vect{H}   \right)^{-1} \vect{P}^{\frac{1}{2}}.
\end{equation}
The heuristic beamforming in \eqref{eq:beamforming-RZF} is known as \emph{regularized zero-forcing beamforming} \cite{Peel2005a} since the identity matrix acts as a regularization of the ZFBF in \eqref{eq:beamforming-highSNR}. Regularization is a common way to achieve numerical stability and robustness to channel uncertainty. Since there is only a single parameter $\lambda$ in regularized ZFBF, it can be optimized for a certain transmission scenario by conventional line search.

The sum property $\sum_{i=1}^{K} \lambda_i = P$ suggests that we set the parameter in regularized ZFBF equal to the average transmit power: $\lambda = \frac{P}{K}$. This parameter choice has a simple interpretation, because the corresponding beamforming directions $( \vect{I}_N +  \sum_{i=1}^{K} \frac{P}{\sigma^2 K} \vect{h}_i \vect{h}_i^H )^{-1} \vect{h}_k$ are the ones that maximize the ratio of the desired signal power to the noise power plus the interference power caused to other users; in other words,
\begin{equation} \label{eq:SLNR-expression}
\argmax{\tilde{\vect{w}}_k : \, \| \tilde{\vect{w}}_k \|^2 = 1 } \quad \frac{ \frac{P}{K \sigma^2} | \vect{h}_k^H \tilde{\vect{w}}_k|^2 }{ \fracSum{i \neq k} \frac{P}{K \sigma^2} | \vect{h}_i^H \tilde{\vect{w}}_k|^2 + 1 } =
\frac{\Big( \vect{I}_N +  \fracSumtwo{i=1}{K} \frac{P}{K \sigma^2 } \vect{h}_i \vect{h}_i^H \Big)^{-1} \vect{h}_k }{ \bigg\| \Big( \vect{I}_N +  \fracSumtwo{i=1}{K} \frac{P}{K \sigma^2} \vect{h}_i \vect{h}_i^H \Big)^{-1} \vect{h}_k  \bigg\|}.
\end{equation}
This heuristic performance metric is identical to maximization of the uplink SINR in \eqref{eq:SINRs_uplink} for equal uplink powers $q_i = \frac{P}{K}$. Hence, \eqref{eq:SLNR-expression} is solved, similar to \eqref{eq:uplink-problem}, as a generalized Rayleigh quotient. The idea of maximizing the metric in \eqref{eq:SLNR-expression} has been proposed independently by many authors and the resulting beamforming has received many different names. The earliest work might be \cite{Zetterberg1995a} from 1995, where the authors suggested beamforming ``such that the quotient of the mean power of the desired contribution to the undesired contributions is maximized''. Due to the relationship to receive beamforming, this scheme is also known as \emph{transmit Wiener filter} \cite{Joham2005a}, \emph{signal-to-leakage-and-noise ratio beamforming} \cite{Sadek2007a}, \emph{transmit MMSE beamforming}, and \emph{virtual SINR beamforming}; see Remark 3.2 in \cite{Bjornson2013d} for a further historical background.

\begin{figure}
         \centering
         \begin{subfigure}[b]{0.5\textwidth}
                 \includegraphics[width=\textwidth]{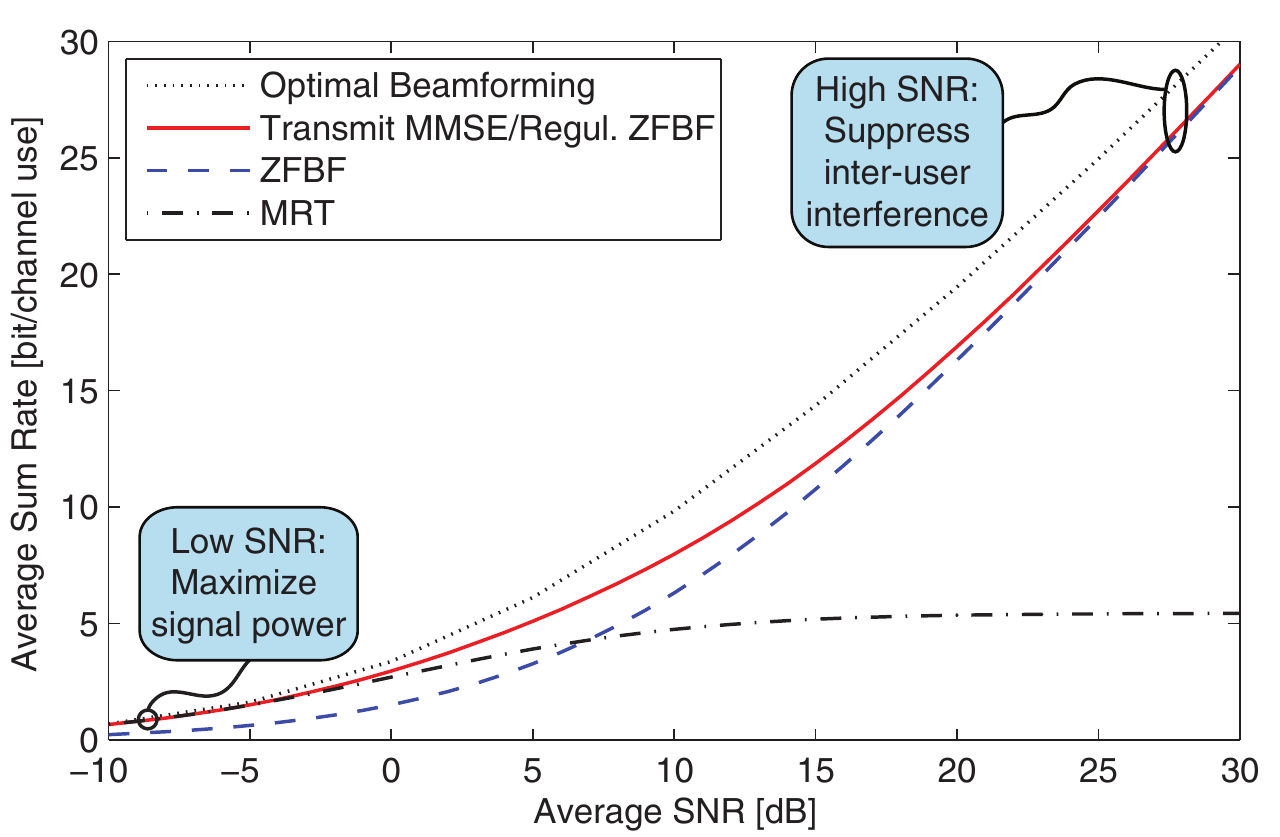} \vskip-1mm
                 \caption{$N=4$ antennas}
                 \label{fig:4antennas}
         \end{subfigure}%
         \begin{subfigure}[b]{0.5\textwidth}
                 \includegraphics[width=\textwidth]{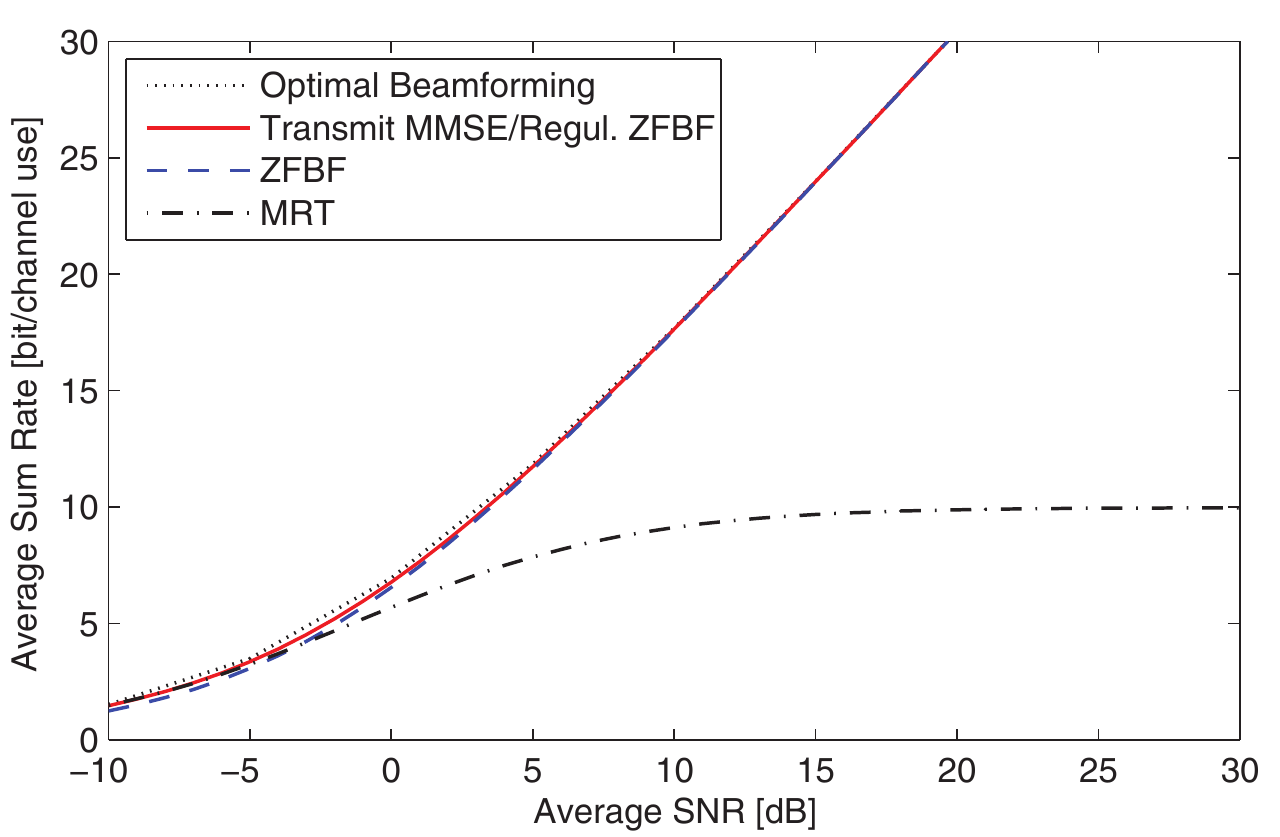} \vskip-1mm
                 \caption{$N=12$ antennas}
                 \label{fig:12antennas}
         \end{subfigure} \vskip-1mm
         \caption{Average sum rate for $K=4$ users as a function of the average SNR. Heuristic beamforming can perform closely to the optimal beamforming, particularly when there are many more antennas than users. Transmit MMSE/regularized ZFBF always performs better, or equally well, as MRT and ZFBF.}\label{fig:simulationexample} \vskip-1mm
\end{figure}

The heuristic beamforming direction in \eqref{eq:SLNR-expression} is truly optimal only in special cases. For example, consider a symmetric scenario where the channels are equally strong and have well separated directivity, while the utility function in \eqref{eq:generalized-problem} is symmetric with respect to $\mathrm{SINR}_1, \ldots, \mathrm{SINR}_K$. It then makes sense to let the $\lambda$-parameters be symmetric as well, which implies $\lambda_k = \frac{P}{K}$ for all $k$ since $\sum_{i=1}^{K} \lambda_i = P$. In other words, the reason that the transmit MMSE beamforming performs well is that it satisfies the optimal beamforming structure---at least in symmetric scenarios. In general, we need all the $K$ degrees of freedom provided by $\lambda_1,\ldots,\lambda_K$ to find the optimal beamforming, because the single parameter in regularized ZFBF does not provide enough degrees of freedom to manage asymmetric user channel conditions and utility functions.

The properties of MRT, ZFBF, and transmit MMSE beamforming are illustrated by simulation in Figure \ref{fig:simulationexample}. We consider $K=4$ users and \eqref{eq:generalized-problem} with the sum rate as utility function: $f(\mathrm{SINR}_1, \ldots, \mathrm{SINR}_4) = \sum_{k=1}^{4} \log_2(1+ \mathrm{SINR}_k) $. The simulation results are averaged over random circularly symmetric complex Gaussian channel realizations, $\vect{h}_k \sim \mathcal{CN}(\vect{0},\vect{I}_N)$, and the SNR is measured as $\frac{P}{\sigma^2}$. The optimal beamforming is computed by the branch-reduce-and-bound algorithm in \cite{Bjornson2013d} whose computational complexity grows exponentially with $K$. This huge complexity stands in contrast to the closed-form heuristic beamforming directions which are combined with a closed-form power allocation scheme from \cite[Theorem 3.16]{Bjornson2013d}.

Figure \ref{fig:simulationexample} shows the simulation results for (a) $N=4$ and (b) $N=12$ transmit antennas. In the former case, we observe that MRT is near-optimal at low SNRs, while ZFBF is asymptotically optimal at high-SNRs. Transmit MMSE beamforming is a more versatile scheme that combines the respective asymptotic properties of MRT and ZFBF with good performance at intermediate SNRs. However, there is still a significant gap to the optimal solution, which is only bridged by fine-tuning the $K=4$ parameters $\lambda_1,\ldots,\lambda_4$ (with an exponential complexity in $K$).
In the case of $N=12$, there are many more antennas than users, which makes the need for fine-tuning much smaller; transmit MMSE beamforming is near-optimal in the entire SNR range, which is an important observation for systems with very large antenna arrays \cite{Rusek2013a} (these systems are often referred to as massive MIMO (multiple-input, multiple-output)). Figure \ref{fig:simulationexample} was generated using Matlab and the code is available for download: \url{https://github.com/emilbjornson/optimal-beamforming}

\section*{Extensions}

Next, we briefly describe extensions to scenarios with multiple BSs and practical power constraints.

\subsection*{Multiple Cooperating Base Stations}

Suppose the $N$ transmit antennas are distributed over multiple cooperating BSs. A key difference from \eqref{eq:generalized-problem} is that only a subset of the BSs transmits to each user, which has the advantage of not having to distribute all users' data to all BSs. This is equivalent to letting only a subset of the antennas transmit to each user. We describe the association by a diagonal matrix $\vect{D}_k = \diag(d_{k,1},\ldots,d_{k,N})$, where $d_{k,n}=1$ if antenna $n$ transmits to user $k$ and $d_{k,n}=0$ otherwise \cite{Bjornson2013d}. The effective beamforming vector of user $k$ is $\vect{D}_k \vect{w}_k$ instead of $\vect{w}_k$. By plugging this into the derivations, the optimal beamforming in \eqref{eq:beamforming-solution2} becomes
\begin{equation} \label{eq:beamforming-solution-cooperating-BSs}
\vect{D}_k \vect{w}_k^\star = \sqrt{p_k} \frac{\Big( \vect{I}_N +  \fracSumtwo{i=1}{K} \frac{\lambda_i}{\sigma^2} \vect{D}_k^H \vect{h}_i \vect{h}_i^H \vect{D}_k \Big)^{-1} \vect{D}_k^H \vect{h}_k }{ \bigg\| \Big( \vect{I}_N +  \fracSumtwo{i=1}{K} \frac{\lambda_i}{\sigma^2} \vect{D}_k^H \vect{h}_i \vect{h}_i^H \vect{D}_k \Big)^{-1} \vect{D}_k^H \vect{h}_k  \bigg\|} \qquad \mathrm{for} \,\,\, k=1,\ldots,K
\end{equation}
for some positive parameters $\lambda_1,\ldots,\lambda_K$. The balancing between high signal power and low interference leakage now takes place only among the antennas that actually transmit to the particular user.

\subsection*{General Power and Shaping Constraints}

Practical systems are constrained not only in terms of the total transmit power, as in \eqref{eq:original-problem} and \eqref{eq:generalized-problem}, but also in other respects; for example, maximal per-antenna power, limited power per BS, regulatory limits on the equivalent isotropic radiated power (EIRP), and interference suppression toward other systems. Such constraints can be well-described by having $L$ quadratic constraints of the form
\begin{equation} \label{eq:power-constraints}
\sum_{k=1}^{K} \vect{w}_k^H \vect{Q}_{\ell,k} \vect{w}_k \leq P_{\ell} \qquad \mathrm{for} \,\,\, \ell=1,\ldots,L
\end{equation}
where the positive semi-definite weighting matrix $\vect{Q}_{\ell,k} \in \mathbb{C}^{N \times N}$ describes a subspace where the power is limited by a constant $P_{\ell} \geq 0$. The total power constraint in \eqref{eq:generalized-problem} is given by $L=1$ and $\vect{Q}_{1,k}= \vect{I}_N$ for all $k$, while per-antenna constraints are given by $L=N$ and $\vect{Q}_{\ell,k}$ being non-zero only at the $\ell$th diagonal element. The weighting matrices are user-specific and can be used for precise interference shaping; for example, the interference leakage at user $i$ is limited to $P_{\ell}$ if we set $\vect{Q}_{\ell,k} = \vect{h}_i \vect{h}_i^H$ for $k \neq i$ and $\vect{Q}_{\ell,i} = \vect{0}$.

If the power constraints are plugged into \eqref{eq:generalized-problem}, it is proved in \cite{Yu2007a,Bjornson2013d} that the optimal beamforming is
\begin{equation}  \label{eq:beamforming-solution_matrixform-constraints}
\vect{W}^\star  = \left( \sum_{\ell=1}^{L} \mu_{\ell} \vect{Q}_{\ell,k} + \frac{1}{\sigma^2} \vect{H} \vect{\Lambda} \vect{H}^H \right)^{-1} \vect{H} \vect{P}^{\frac{1}{2}}
\end{equation}
where the $L$ new parameters $\mu_1,\ldots,\mu_L \geq 0$ describe the importance of shaping the beamforming to each power constraint; if $\mu_{\ell}$ is large, very little power is transmitted into the subspace of $\vect{Q}_{\ell,k}$. On the contrary, inactive power constraints have $\mu_{\ell}=0$ and, thus, have no impact on the optimal beamforming. One can show that the parameters satisfy $\sum_{i=1}^{K} \lambda_i = P_{\max}$ and $\sum_{\ell=1}^{L} P_{\ell} \mu_{\ell} = P_{\max}$ where $P_{\max} = \max_{\ell} P_{\ell}$ \cite{Bjornson2013d}.

\section*{Lessons Learned and Future Avenues}

It is difficult to compute the optimal multiuser transmit beamforming, but the solution has a simple and intuitive structure with only one design parameter per user. This fundamental property has enabled many researchers to propose  heuristic beamforming schemes---there are many names for essentially the same simple scheme. The optimal beamforming maximizes the received signal powers at low SNRs, minimizes the interference leakage at high SNRs, and balances between these conflicting goals at intermediate SNRs.

The optimal beamforming structure can be extended to practical multi-cell scenarios, as briefly described in this lecture. Alternative beamforming parameterizations based on local channel state information (CSI) or transceiver hardware impairments can be found in \cite{Bjornson2013d}. Some open problems in this field are the robustness to imperfect CSI, multi-stream beamforming to multi-antenna users, multi-casting where each signal is intended for a group of users, and adaptive $\lambda$-parameter selection based on the utility function.

\section*{Acknowledgment}

This work was supported by the International Postdoc Grant 2012-228 from the Swedish Research Council and by the European Research Council under the European Community's Seventh Framework Programme (FP7/2007–2013)/ERC Grant 228044.

\section*{Authors}

\textit{Emil Bj\"ornson} (emil.bjornson@liu.se) is an assistant professor at the Dept.~of Electrical Engineering, Link\"oping University, Sweden. He is also a joint postdoctoral researcher at Sup\'elec, France, and at the KTH Royal Institute of Technology, Sweden.

\textit{Mats Bengtsson} (mats.bengtsson@ee.kth.se) is an associate professor at the Dept.~of Signal Processing, KTH Royal Institute of Technology, Sweden.

\textit{Bj\"orn Ottersten} (bjorn.ottersten@ee.kth.se) is a professor at the Dept.~of Signal Processing, KTH Royal Institute of Technology, Sweden, and also the director for the Interdisciplinary Centre for Security, Reliability and Trust, University of Luxembourg, Luxembourg.

\section*{Reproducible Research}

This lecture note has supplementary downloadable material available at \url{https://github.com/emilbjornson/optimal-beamforming}, provided by the authors. The material includes Matlab code that can reproduce all the simulation results. Contact emil.bjornson@liu.se for further questions about this work.

\bibliographystyle{IEEEtran}
\bibliography{IEEEabrv,refs}

\end{document}